%% file: main.tex
\documentclass[10pt,conference,screen]{IEEEtran}
\IEEEoverridecommandlockouts

\input{macros}

\begin{document}

\title{{\toolname: Multi-Fidelity Simulation-based Testing of Automated Driving Systems}
\thanks{This research was partly funded by the Bavarian Ministry of Economic Affairs, Regional Development and Energy and partly supported by AI4SDV, funded by the Chips Joint Undertaking and its members under grant agreement No 101329675, co-funded by the European Union.}
}

\author{
\IEEEauthorblockN{
Lev Sorokin\textsuperscript{1},
Berfin Dinler\textsuperscript{1},
Alessio Gambi\textsuperscript{2,3},
Andrea Stocco\textsuperscript{1,4}
}
\IEEEauthorblockA{
\textsuperscript{1}\textit{Technical University of Munich},
Munich, Germany,
\{lev.sorokin, berfin.dinler, andrea.stocco\}@tum.de
}
\IEEEauthorblockA{
\textsuperscript{2}\textit{AIT - Austrian Institute of Technology},
Vienna, Austria,
alessio.gambi@ait.ac.at
}
\IEEEauthorblockA{
\textsuperscript{3}\textit{The Italian Institute of Artificial Intelligence (AI4I)},
Torino, Italy
alessio.gambi@ai4i.it
}
\IEEEauthorblockA{
\textsuperscript{4}\textit{fortiss GmbH},
Munich, Germany,
stocco@fortiss.org
}
}

\maketitle

\input{content/abstract}


\input{content/intro}
\input{content/background}
\input{content/approach}
\input{content/evaluation}

\input{content/discussion}
\input{content/related}
\input{content/conclusion}



\bibliographystyle{IEEEtran}
\bibliography{papers}

\end{document}

%% file: macros.tex
\usepackage{enumitem}
\usepackage{orcidlink} 
\usepackage{multirow}%
\usepackage{mathrsfs}%
\usepackage{manyfoot}%
\usepackage{array}
\usepackage{amsfonts}
\usepackage{cite}
\usepackage{algorithmic}
\usepackage{graphicx}
\usepackage{textcomp}
\usepackage{xcolor}
\usepackage{listings}
\usepackage{sansmath}
\usepackage{hyperref}
\usepackage{tikz}
\usepackage{tcolorbox}
\tcbuselibrary{skins}
\usepackage{nicefrac}
\usepackage{pbalance}
\shrinkLastPage{74pt}
\usepackage{lipsum}
\usepackage{subcaption}
\usepackage{float}
\usepackage{booktabs}
\usepackage{xspace}
\usepackage{tikz}
\usepackage{adjustbox}
\usepackage{subcaption}
\usepackage{amsmath}
\usepackage{wrapfig}
\usepackage{physics}
\usepackage{tablefootnote}
\usepackage{threeparttable}
\usepackage{caption} 
\usepackage{lineno}
\usepackage[normalem]{ulem}
\usepackage{makecell}
\usepackage{xcolor}
\usepackage{colortbl}
\usepackage{graphicx}
\usepackage{amsmath,amssymb}
\usepackage{theorem}
\usepackage{placeins}
\usepackage{float}

\usepackage[vlined, boxruled, linesnumbered] {algorithm2e}
\SetKwComment{Comment}{$\triangleright$\ }{}
\SetCommentSty{itshape}
\newboolean{showcomments}
\setboolean{showcomments}{true}
\ifthenelse{\boolean{showcomments}}
{
	\definecolor{myyellow}{RGB}{255, 228, 26}
	\definecolor{myblue}{RGB}{50, 50, 220}
	\newcommand{\nb}[2]{
		{\sf
			\fcolorbox{myyellow}{yellow}{\scriptsize\textbf{#1}}%
			$\blacktriangleright$%
			{\color{myblue}\fontsize{9pt}{9pt}\selectfont\textbf{#2}}%
		}%
	}
}
{
	\newcommand{\nb}[2]{}
}

\newcommand\Alessio[1]{\nb{Alessio}{#1}}

\newcommand{\lofi}[0]{LoFi\xspace}
\newcommand{\hifi}[0]{HiFi\xspace}

\newcommand{\toolname}[0]{SimFusion\xspace}

\newcommand{\autoware}[0]{Autoware\xspace}
\newcommand{\frenetix}[0]{FrenetiX\xspace}

\usepackage[normalem]{ulem}

\usepackage{soul}

\hypersetup{
    pdfborder={0 0 0}
}

\newcommand{\head}[1]{\noindent\textbf{#1.}}
\newcommand{\figstep}[1]{\tikz[baseline=(s.base)]\node[circle,fill=black,text=white,font=\scriptsize\bfseries\sffamily,inner sep=0pt,minimum size=3.5mm](s){#1};}



%% file: content/abstract.tex
\begin{abstract}

Simulation is the primary means of validating automated driving systems (ADS).
Testing typically relies on a single simulator, forcing a trade-off between the speed of low-fidelity (LoFi) simulation and the realism of computationally expensive high-fidelity (HiFi) simulation.
We present \toolname, a multi-fidelity search-based testing framework that decides, for every candidate scenario, which simulator should execute it. Rather than approximating the ADS or its fitness, \toolname learns whether LoFi and HiFi simulators are likely to agree on a scenario's outcome: scenarios predicted to agree run in the fast LoFi simulator, while all others are conservatively run in HiFi.
We evaluate \toolname on two ADS stacks, three driving scenarios, and four simulators, against LoFi-only, HiFi-only, and surrogate-assisted testing. Across six runs per configuration, \toolname finds on average $2.1\times$ more HiFi-validated failures than HiFi-only testing and $1.6\times$ more than surrogate-assisted testing, up to $3\times$ and $1.7\times$ faster per failure, while executing $31\%$ fewer HiFi simulations than HiFi-only testing. \toolname also achieves the highest failure coverage in all three scenarios.
An ablation with random simulator selection shows that the gains stem from the learned agreement model rather than from combining two simulators alone.
These results suggest that predicting cross-simulator agreement is an effective alternative to surrogate modeling, as simulation fidelity can serve as an optimization dimension in ADS testing.

\end{abstract}


%% file: content/intro.tex
\section{Introduction}\label{sec:introduction}

Automated Driving Systems (ADS) are primarily validated through simulation before real-world testing~\cite{survey-lei-ma,DBLP:journals/corr/abs-2101-05337}. 
Simulators offer varying trade-offs between computational cost and realism~\cite{DBLP:conf/ivs/OsikowiczMWS26,2026-Lambertenghi-arxiv}. 
High-fidelity (HiFi) simulators, such as CARLA~\cite{Dosovitskiy17} and BeamNG.tech~\cite{beamng}, accurately model vehicle dynamics and perception, but their execution is time-consuming, requiring from tens of seconds to several minutes per scenario in our setting.
Conversely, low-fidelity (LoFi) simulators, such as CommonRoad~\cite{commonroad} and MetaDrive~\cite{metadrive}, abstract these components to enable faster execution at the cost of reduced realism~\cite{DBLP:conf/ivs/SagmeisterKGL24,kaufeld2025multidrive,2024-Lambertenghi-ICST}.

Despite the availability of multiple simulators, simulation fidelity remains a fixed design choice in current testing pipelines. LoFi simulators are typically used for rapid prototyping, whereas HiFi simulators are reserved for system-level validation~\cite{ul4600,2023-Stocco-EMSE}. Likewise, existing search-based testing executes all candidate scenarios in a single simulator, forcing developers to trade exploration throughput for realism~\cite{challengesADSTesting}. 

Some approaches attempt to reduce HiFi executions using surrogate models that predict failures or fitness values~\cite{birchler2023machine,DBLP:conf/icse/Haq0B22,morlot,JodatSpurious23,2026-Naziri-ICSE,2020-Stocco-ICSE,2022-Stocco-ASE,2024-Grewal-ICST}. 
However, these approaches require learning a surrogate of the entire ADS, i.e., a mapping from high-dimensional driving scenarios to complex closed-loop system behavior. Such behavior emerges from the interaction between perception, planning, control, and the environment, making accurate approximation challenging.
Other approaches reduce execution cost by adjusting simulator parameters, such as the frame rate, trading simulation accuracy for efficiency~\cite{DBLP:conf/ivs/OsikowiczMWS26}.

In this paper, we present \toolname, a multi-fidelity search-based testing framework that dynamically selects the simulation fidelity at which each candidate scenario is evaluated. 
Rather than learning failures or the whole ADS behavior, \toolname learns a model of \emph{cross-simulator agreement} between LoFi and HiFi simulators to predict whether both simulators are likely to produce the same test outcome for a given scenario~\cite{matteoMaxibon2024}.
Since search-based testing spends most of its computational budget evaluating non-critical scenarios, identifying when these scenarios can be evaluated reliably in the faster LoFi simulator offers the greatest opportunity to reduce testing cost. \toolname therefore uses the agreement model to guide simulator selection, reserving expensive HiFi execution for scenarios where HiFi simulation is predicted to reach a different safety verdict than LoFi. 



We evaluate \toolname across two ADS stacks, three driving scenarios, and four simulators: CARLA~\cite{Dosovitskiy17} and BeamNG.tech~\cite{beamng} as HiFi simulators, and CommonRoad~\cite{commonroad} together with a ROS-based kinematic simulator as LoFi simulators. 
Across six runs per configuration, \toolname finds on average $2.1\times$ more HiFi-validated failures than HiFi-only testing and $1.6\times$ more than surrogate-assisted testing, finds them up to $3\times$ and $1.7\times$ faster, achieves the highest failure coverage in all three scenarios, and executes $31\%$ fewer HiFi simulations than HiFi-only testing. 

Beyond these empirical improvements, our study provides two key insights. First, our results suggest that predicting when HiFi simulation is worthwhile is a more tractable learning problem than approximating fitness functions or entire ADS behaviors, and that it makes simulator orchestration more effective than surrogate-assisted testing in our case studies. Second, treating simulator fidelity as a runtime decision yields better failure discovery than committing to a fixed simulator.

The main contributions of this paper are:

\begin{description}[noitemsep]
\item[\toolname.]
A multi-fidelity search-based testing framework that dynamically orchestrates low- and high-fidelity simulators by learning cross-simulator agreement~\cite{replication-package}.

\item[Empirical evaluation.]
A comprehensive evaluation across two ADS stacks, three driving scenarios, four simulators, that comprises more than $27{,}000$ test executions.

\item[Findings.]
Empirical evidence that using cross-simulation agreement to guide simulator orchestration leads to the discovery of substantially more failures than both fixed-fidelity and surrogate-assisted approaches while improving testing efficiency and reducing HiFi executions.
\end{description}

%% file: content/background.tex
\section{Background}\label{sec:background}

\begin{figure}[t]
    \centering
    \begin{subfigure}[t]{0.488\linewidth}
        \centering
        \includegraphics[width=\linewidth]{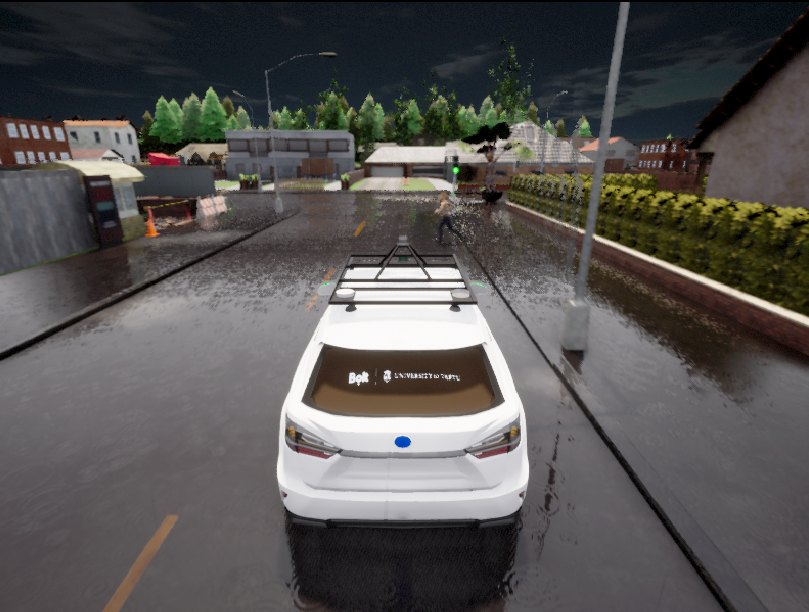}
        \caption{\hifi.}
        \label{fig:fig1}
    \end{subfigure}
    \hfill
    \begin{subfigure}[t]{0.465\linewidth}
        \centering
        \includegraphics[width=\linewidth]{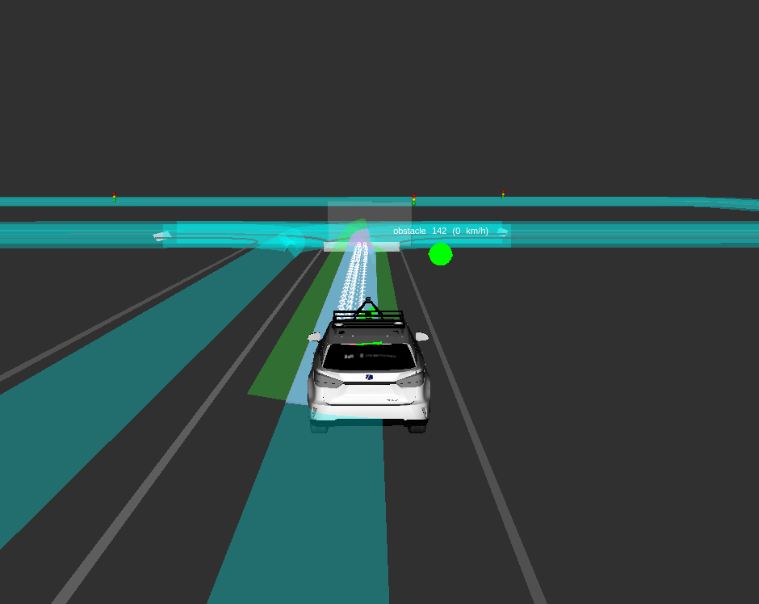}
        \caption{\lofi.}
        \label{fig:fig2}
    \end{subfigure}
\caption{Comparison of \hifi (CARLA) and \lofi simulations of the same automated emergency braking scenario.}
    \label{fig:combined}
\end{figure}

\subsection{Simulation Fidelity in ADS Testing}

In ADS testing, \emph{simulation fidelity} denotes how accurately a simulator reproduces sensing, vehicle dynamics, and interactions with the environment~\cite{2023-Stocco-TSE}. Simulators execute \emph{driving scenarios}, which specify the initial state and behavior of the ego vehicle (i.e., the ADS under test), the surrounding traffic participants, and the environment.

High-fidelity simulators, such as CARLA~\cite{Dosovitskiy17} and BeamNG.tech~\cite{beamng}, explicitly model sensing, vehicle dynamics, and environmental interactions using detailed physics and photorealistic rendering.
These simulators generate realistic sensor inputs (e.g., camera images or LiDAR point clouds) that are processed by the ADS perception stack, yielding realistic yet computationally expensive simulations.

Low-fidelity simulators simplify sensing, dynamics, or both to reduce execution time. They typically replace sensor simulation with ground-truth state information and approximate vehicle motion using kinematic models. Examples include CommonRoad~\cite{commonroad}, MetaDrive~\cite{metadrive}, and esmini~\cite{esmini}. These abstractions substantially accelerate simulation but may miss failures that depend on realistic perception or physics.

Figure~\ref{fig:combined} compares the \hifi and \lofi implementations of the automated emergency braking (AEB) scenario used throughout this paper. \hifi relies on LiDAR-based perception and rigid-body physics, whereas \lofi uses ground-truth perception and a kinematic bicycle model. In our experiments, these abstractions make a \lofi test between $1.8\times$ (AEB) and $4\times$ (FrenetiX) faster than its \hifi counterpart, at the cost of potentially changing whether a scenario is classified as safe or unsafe.

\subsection{Simulator Orchestration Problem}

Prior work orchestrates multiple \hifi simulators to expose simulator-specific failures and filter out spurious ones~\cite{matteoMaxibon2024,multisim2026sorokin}, but running several \hifi simulators multiplies the cost of testing. We instead orchestrate across fidelity levels, running most scenarios in a cheap \lofi simulator and invoking \hifi only when necessary.

The difficulty is that the same scenario can yield different outcomes in different simulators~\cite{digitalTwinAreNotMonozygotic,2025-Lambertenghi-ASE}. A scenario deemed safe in \lofi may fail in \hifi, and a \lofi failure may not reproduce in \hifi; in both cases, the search follows a misleading signal, discarding promising scenarios or chasing spurious ones. Orchestration therefore requires knowing, before execution, when two simulators are likely to reach the same verdict and when additional fidelity is needed.

%% file: content/approach.tex
\section{Search-Based Scenario Generation Driven by Simulation Orchestration}\label{sec:approach}

\begin{figure*}
    \centering
    \includegraphics[width=0.95\linewidth]{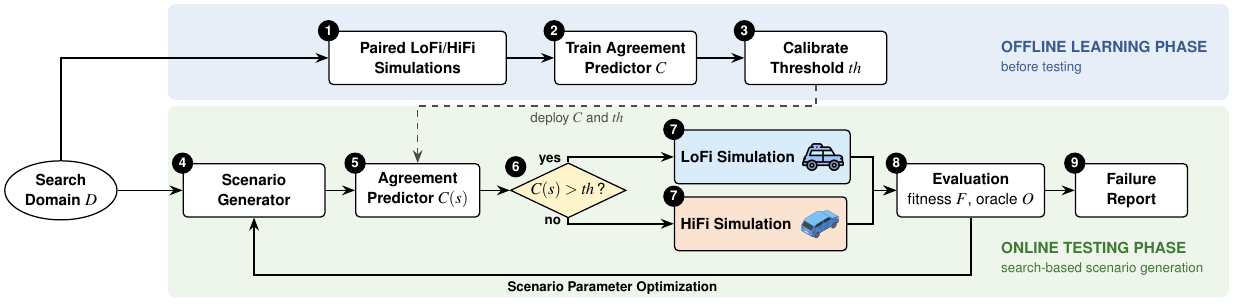}
    \caption{Overview of \toolname.}
    \label{fig:overview-simfusion}
\end{figure*}

\toolname is a multi-fidelity search-based testing framework that decides, for every generated scenario, whether it must be executed in the HiFi simulator or whether the faster LoFi simulator suffices. 
\toolname takes as input the ego-vehicle (the $ADS$ under test), the search domain $D$ (which defines the valid ranges of the scenario parameters), a fitness function $F$ (which quantifies how critical a scenario is), and an oracle $O$ (which distinguishes passing from failing scenarios). 
In addition, it requires a pair of \lofi and \hifi simulators that can execute the same scenarios against the ADS and evaluate them with the same $F$ and $O$, albeit at different fidelity levels.

\autoref{fig:overview-simfusion} gives an overview of \toolname, which operates in two phases. 
In the \emph{offline learning phase}, \toolname learns from paired \lofi/\hifi executions in which regions of the search domain both simulators agree on the test outcome. 
In the \emph{online testing phase}, it uses this knowledge to orchestrate the two simulators during search-based scenario generation, as outlined in \autoref{algo:algo1}. 
In the following, we first define simulator agreement and then describe both phases along the steps in \autoref{fig:overview-simfusion}.

\subsection{Simulator Agreement}
\label{sec:agreement}

Executing a scenario $s$ in a simulator $S \in \{\textit{LoFi}, \textit{HiFi}\}$ yields an oracle outcome $o \in \{\textit{failure}, \textit{non\mbox{-}failure}\}$, indicating whether the ADS violates a system requirement. 
In this paper, we focus on safety violations such as collisions, violations of the minimum safety distance, and critical time-to-collision values~\cite{LIU2021103037}. 
Let $o_{LoFi}(s)$ and $o_{HiFi}(s)$ denote the outcomes of executing $s$ in \lofi and \hifi, respectively. 
The simulators \emph{agree} on $s$ if $o_{LoFi}(s)=o_{HiFi}(s)$, and \emph{disagree} otherwise. 
A disagreement occurs in two forms:

\begin{description}[leftmargin=1.8em]
\item[\textbf{Case 1}] \lofi reports a failure, whereas \hifi does not.
\item[\textbf{Case 2}] \hifi reports a failure, whereas \lofi does not.
\end{description}

From a search-based testing perspective, the two cases have different consequences. 
Case~1 wastes search effort by following spurious \lofi failures, whereas Case~2 is more critical, as genuine \hifi failures are missed. 
\toolname therefore favors \hifi whenever a disagreement is likely.

\subsection{Offline Learning Phase}
\label{sec:offline}

The offline phase (upper part of \autoref{fig:overview-simfusion}) is executed once before testing and prepares the simulator orchestration.

In the first step~\figstep{1}, \toolname samples a set of scenarios from the search domain $D$, executes each of them in both \lofi and \hifi, and labels it as \emph{agree} or \emph{disagree} by comparing $o_{LoFi}(s)$ and $o_{HiFi}(s)$.

In the second step~\figstep{2}, these labeled pairs are used to train a binary classifier $C$, which maps the parameters of a scenario $s$ to the probability $C(s)$ that \lofi and \hifi agree on its outcome.
Unlike surrogate-assisted approaches, $C$ does not need to approximate the behavior of the ADS or predict continuous fitness values; it only learns whether increased fidelity is likely to change the oracle verdict.
This is a substantially simpler learning problem that still provides the information required for orchestration.
\toolname is agnostic to the classifier type; in our evaluation, we compare several common classifiers (Section~\ref{sec:evaluation}).

In step~\figstep{3}, a confidence threshold $th$ is calibrated, which determines how confident $C$ must be in an agreement before the \lofi result is trusted.
A higher $th$ sends more scenarios to \hifi, trading efficiency for a lower risk of missing failures (Case~2).
In our evaluation, we select $th$ using the Youden index~\cite{Youden}.

\subsection{Online Testing Phase}
\label{sec:online}

The online phase (lower part of \autoref{fig:overview-simfusion}) integrates the trained predictor $C$ and threshold $th$ into a search-based testing loop that iteratively generates, evaluates, and evolves scenarios (\autoref{algo:algo1}).

\begin{algorithm}[t]
\DontPrintSemicolon
\footnotesize

\SetKwInOut{Input}{Input}
\SetKwInOut{Output}{Output}

\Input{
$(ADS, D, F, O)$: SBST problem instance \newline
$S_{lofi}$: LoFi simulator \newline
$S_{hifi}$: HiFi simulator \newline
$C$: agreement classifier \newline
$th$: classifier confidence threshold
}

\Output{$Failures$: set of failure-inducing test inputs}

\BlankLine

$T \gets \emptyset$\;
$P \gets \emptyset$ \tcp*{$T$: archive of evaluated scenarios, $P$: current population}

$P \gets \textsc{sampling}(ADS,D)$\;

$P \gets \textsc{runOrchestrator}(P,S_{lofi},S_{hifi},C,th,F,ADS)$\;

\While{$budgetAvailable$}{

$Par \gets \textsc{tournamentSelection}(P)$\;

$P \gets \textsc{crossoverMutate}(Par)$\;

$P \gets \textsc{runOrchestrator}(P,S_{lofi},S_{hifi},C,th,F,ADS)$\;

$T \gets T \cup P$\;

$P \gets \textsc{survival}(P,F)$\;

%
}

$Failures \gets \{x \in T \mid O(x)=1\}$\;

\Return{$Failures$}

\caption{\toolname}
\label{algo:algo1}
\end{algorithm}

In step~\figstep{4}, the scenario generator creates the scenarios to be executed.
In the first iteration, it samples an initial population from $D$ using Latin Hypercube Sampling~\cite{LHS} (\autoref{algo:algo1}, line~3).
In subsequent iterations, it evolves the population using evolutionary operators~\cite{Sorokin2024} (lines~6--7):
tournament selection picks parent scenarios, favoring higher fitness while maintaining diversity; 
Simulated Binary Crossover~\cite{deb2007self_adaptive_sbx} recombines the parameters of two parents (e.g., ego speed and pedestrian distance); 
and Polynomial Mutation~\cite{deb2007self_adaptive_sbx} perturbs each parameter with probability $\mu_{mut}$ to explore nearby regions of the search space. 
Scenarios whose parameters match previously generated ones are discarded and regenerated to avoid redundant simulations.

Next, \toolname decides for each generated scenario $s$ which simulator executes it.
In step~\figstep{5}, the trained classifier computes $C(s)$, and in step~\figstep{6}, this value is compared against the threshold.
If $C(s) > th$, the simulators are expected to agree and $s$ is executed in the faster \lofi simulator; otherwise, it is conservatively executed in \hifi (step~\figstep{7}, lines~4 and~8 of \autoref{algo:algo1}).

In the evaluation step~\figstep{8}, the trace produced by the selected simulator is used to compute both the fitness value $F(s)$ and the oracle verdict $O(s)$.
Unlike the binary oracle, the fitness function is a continuous objective that quantifies how close $s$ is to violating the target requirement, e.g., the minimum distance between the ego vehicle and other traffic participants.
\toolname is agnostic to the specific choice of $F$.

The fitness values then close the search loop (\emph{Scenario Parameter Optimization}): evaluated scenarios are added to the archive $T$, and survival (lines~9--10) determine which scenarios form the next population, which the generator evolves again in step~\figstep{4}.
In this way, the search is steered toward increasingly critical scenarios.

When the testing budget is exhausted, \toolname returns all oracle-violating scenarios in the archive (step~\figstep{9}), i.e., $\{s \in T \mid O(s)=1\}$, as failure-inducing scenarios (lines~11--12).

\subsection{Implementation}
\label{sec:implementation}

We implemented \toolname in Python on top of OpenSBT~\cite{sorokin2023opensbt}, a modular framework for search-based testing of ADS that has been used in similar research~\cite{multisim2026sorokin,Sorokin2024,sorokin24svm}. The simulator orchestrator is integrated into the evaluation step of the search loop, so that the generator and the survival operator, and the fitness and oracle computation are shared by \toolname and all baselines. The implementation is publicly available~\cite{replication-package}.

For \autoware, CARLA, the ROS-based \lofi simulator, and Autoware~Mini are deployed in Docker containers. Scenarios in CARLA are executed via CarlaRunner~\cite{sorokin2023opensbt}, whereas \lofi simulations are triggered through ROS message exchange~\cite{autoware_mini}. For \frenetix, BeamNG.tech runs on Windows, while CommonRoad and FrenetiX are deployed in WSL due to dependency constraints, and scenario execution across the two simulators is kept consistent through MultiDrive~\cite{kaufeld2025multidrive}. To ensure the robustness of long-running experiments, we implement a simulator restart mechanism that recovers from occasional execution interruptions.

%% file: content/evaluation.tex
\section{Evaluation}
\label{sec:evaluation}

\subsection{Research Questions}

\noindent\textbf{RQ\textsubscript{1} (agreement).}
\textit{How consistently do \lofi and \hifi simulations agree on the outcome of a scenario?}
The aim is to investigate to what extent low-fidelity and high-fidelity simulators agree on the outcome of the same driving scenarios. This question serves as a prerequisite for our approach. It assesses whether simulator agreement is sufficiently systematic to enable reliable simulator orchestration.

\noindent\textbf{RQ\textsubscript{2} (effectiveness).}
\textit{How effective is \toolname at detecting failures compared to single-simulation and surrogate-based approaches?}
This research question evaluates whether dynamic orchestration of \lofi and \hifi simulations improves failure detection compared to executing scenarios in a single simulator or relying on surrogate models.


\noindent\textbf{RQ\textsubscript{3} (efficiency).}
\textit{How efficient is \toolname compared to single-simulation and surrogate-based approaches?}
This research question evaluates whether \toolname reduces the computational cost of testing while maintaining effective failure detection through selective \hifi execution.


\noindent\textbf{RQ\textsubscript{4} (failure diversity).}
\textit{How diverse are the failures identified by \toolname compared to failures obtained through single-fidelity simulation approaches?}
This research question evaluates the diversity of the failure-inducing scenarios identified by \toolname and compares their distribution across the scenario space with those obtained using single-simulation and surrogate-based approaches.


\subsection{Objects of Studies}
\label{sec:objects}

To evaluate the generality of our approach, we consider two realistic ADS stacks and three driving scenarios. Both ADS stacks implement contemporary architectural paradigms and perception/planning pipelines, are developed in different simulation ecosystems, and deployed on real vehicles. 

\subsubsection{Autoware~Mini}
Autoware Mini~\cite{autoware_mini} is a Python port of the autonomous driving stack developed by the Autoware Foundation~\cite{autoware-foundation-github}.
The system is a fully-fledged ADS which includes obstacle detection, global and local planning, control, and a camera-based traffic-light detector.
Autoware~Mini is deployed on an experimental autonomous vehicle (a Lexus RX450h) at the University of Tartu~\cite{adlvehicle}.
As \hifi simulation, we use CARLA~\cite{Dosovitskiy17}.
As \lofi simulator we use the ROS message exchange bundled with Autoware Mini. 
Compared to CARLA, which uses LiDAR perception and rigid-body physics, the \lofi simulator provides ground-truth data on obstacles and traffic lights. Furthermore, it uses the bicycle model to simulate vehicle dynamics.
To ensure that outcome differences are attributable to simulator fidelity rather than mismatched scenario executions, we executed a pilot study to assess trace alignment across simulator executions using distance metrics (i.e., Euclidean, DTW~\cite{DTW}, and Hausdorff~\cite{huettenlocher1993hausdorf}) as detailed in our replication package~\cite{replication-package}.

We evaluate Autoware~Mini on two abstract scenarios, widely used in prior work~\cite{Abdessalem-ICSE18}: (a)~an automated emergency-braking (AEB) scenario in which the ego vehicle drives on a straight road segment while a pedestrian crosses the road in front of it; (b)~a cut-in scenario (CUT-IN) in which a slower vehicle traveling in front of the ego-vehicle in the left lane performs a cut-in maneuver into the ego vehicle's lane.
Scenarios take place on the \texttt{Town1} and \texttt{Town4} maps, respectively. 
The AEB scenario is parameterized by the ego vehicle's target velocity $EgoSpeed$, the pedestrian trigger distance $PedDist$, and the pedestrian velocity $PedSpeed$. Parameter ranges are derived from prior studies and preliminary experiments~\cite{nejati2023reflections}. 
The CUT-IN scenario is parametrized by the ego vehicle's target velocity $EgoSpeed$, the adversary vehicle triggering distance $DistStartCutin$, the cut-in duration $LengthCutin$, and the speed difference between the vehicles $SpeedDiff$.

\begin{table}[t]
\caption{FrenetiX scenario variables and ranges.}
\label{tab:frenetix-ranges}
\centering
\resizebox{0.9\linewidth}{!}{
\begin{tabular}{l l l}
\toprule
Parameter & Range & Description \\
\midrule
goal\_center\_x     & [150, 160]   & Goal longitudinal position \\
goal\_center\_y     & [0, 0.5]     & Goal lateral position \\
goal\_velocity\_min & [10, 15]     & Minimum goal velocity m/s \\
goal\_velocity\_max & [20, 25]     & Maximum goal velocity m/s \\
num\_npc            & [2, 4]       & Number of NPC vehicles \\
lane\_count         & [2, 5]       & Number of lanes \\
npc\_x0             & [9, 12]      & Initial NPC position (x) \\
npc\_v\_max         & [15, 20]     & Maximum NPC velocity m/s \\
npc\_seed           & [0, 99]      & NPC behavior seed \\
cutin\_tx           & [0.5, 1.0]   & Scalar cut-in duration \\
cutin\_delay        & [0, 0.5]     & Scalar cut-in activation delay \\
\bottomrule
\end{tabular}}
\end{table}

\subsubsection{FrenetiX}
FrenetiX~\cite{frenetix} is an open-source motion planning framework implementing an efficient sampling-based trajectory planning in Python and C++.
FrenetiX is deployed in the EDGAR research vehicle (a Volkswagen T7 Multivan) at the Technical University of Munich~\cite{karle2024edgarautonomousdrivingresearch}.
As \hifi simulator, we use BeamNG.tech~\cite{beamng}, which implements photo-realistic environments and soft-body physics, whereas as \lofi simulator, we use CommonRoad~\cite{commonroad}, which implements a lightweight 2D simulation.
We use the default simulator configuration and a highway template from CommonRoad~\cite{commonroad} and ensure consistent scenario execution using MultiDrive~\cite{kaufeld2025multidrive}.

We evaluate FrenetiX in a highway scenario involving a lane-change maneuver on a straight, multi-lane road with surrounding traffic~\cite{kaufeld2025multidrive}. The scenario is parameterized by $11$ variables that control the target position and velocity of the ego vehicle, the number of lanes and vehicles, and the cut-in maneuvers of surrounding traffic (\autoref{tab:frenetix-ranges}).
%

\subsection{\toolname Configurations}

To generate critical scenarios, i.e., scenarios in which the ego-vehicle collides at high speed, we configure \toolname based on preliminary runs and best practices (see~\autoref{tab:experiment_configuration}).
The search loop runs for $3.0$ hours and evolves a population of $20$ individuals in the AEB scenario, while it runs for $2.5$ hours and evolves a population of $15$ individuals in the CUT-IN scenario.
Mutations and crossover trigger with probabilities $0.2$ and $0.8$ in both scenarios.
%

\head{Autoware~Mini}
In the AEB scenario, the test oracle flags a failure if the ego-vehicle is too close to the pedestrian crossing in front of it ($F_{dist} < 1$) while the ego-vehicle is moving ($F_v > 0$).
Similarly, in the CUT-IN scenario, it flags a failure when the TTC falls below $0.5$~s~\cite{Quante2021HumanPerformance}.

\head{FrenetiX}
We configure \toolname to 
minimize the distance between the ego-vehicle and the other vehicles ($F_{dist}$) while maximizing the distance of the ego-vehicle from the 
goal area it must reach ($F_{goal}$).
To account for longer simulations in BeamNG.tech, \toolname evolves a population of $10$ individuals for $6$ hours. Mutation and crossover trigger with the same probabilities used for Autoware Mini.
%
The test oracle flags a failure if $F_{dist} < 0.5m$ or $F_{goal} > 2m$, indicating unsafe proximity or failure to reach the goal.

\begin{table}[t]
\centering
\caption{Algorithm configuration for the approaches under test \lofi, \hifi, and \toolname, summarized by case study.}
\label{tab:experiment_configuration}
\resizebox{0.97\linewidth}{!}{%
\begin{tabular}{p{2cm} l l}
\toprule
\textbf{Parameter} & \textbf{\autoware (AEB/CUT-IN)} & \textbf{\frenetix} \\
\midrule
Simulators & Carla\cite{Dosovitskiy17}, SimpleSim\cite{autoware_mini} & BeamNG~\cite{beamng}, CommonRoad~\cite{commonroad} \\
Test Generator & NSGA-II & NSGA-II \\
Population size & 20/15 & 10 \\
Time budget & 3/2.5 hours & 6 hours \\
Mutation rate & 0.33/0.25 & 0.09 \\
Crossover rate & 0.6 & 0.6 \\
Fitness functions & 2 & 2 \\
Search variables & 3/4 & 10 \\
\bottomrule
\end{tabular}%
}
\end{table}

\subsection{Baselines}

We compare \toolname against two baselines: single-fidelity testing and surrogate-assisted testing.
We instantiate single-fidelity in two versions: single-fidelity testing using only \hifi and single-fidelity testing using only \lofi. 

The surrogate-assisted baseline follows the approach proposed by Jodat et al.~\cite{JodatSpurious23}. It leverages a regression model to predict the fitness values of newly generated scenarios before executing them. The surrogate model is trained on labeled execution data from \hifi simulations. Specifically, the surrogate-assisted baseline uses the prediction uncertainty to determine whether the prediction interval, estimated from the validation-set mean absolute error, lies entirely within either the pass or fail region. 
If the interval lies entirely within a single region, the prediction is considered reliable, and the surrogate result is used without executing the \hifi simulator; otherwise, the scenario is evaluated with \hifi.

To ensure a fair comparison, all approaches use the same search-based testing framework and optimization as \toolname, differing only in how candidate scenarios are evaluated.

\begin{table}[t]
\centering
\caption{Features used for diversity analysis.}
\begin{tabular}{p{2.5cm}ccc}
\toprule
\textbf{Feature} &
\textbf{FrenetiX} &
\textbf{Autoware AEB} &
\textbf{Autoware CUT-IN} \\
\midrule
min\_TTC                          & \checkmark & \checkmark & \checkmark \\
min\_dist                         & \checkmark & \checkmark & \checkmark \\
top2\_min\_dist                   & \checkmark & - & - \\
num\_neighbors\_within\_5m         & \checkmark & - & - \\
time\_with\_neighbors\_5m          & \checkmark & - & \checkmark \\
min\_lateral\_gap                 & \checkmark & - & \checkmark \\
min\_longit\_gap            & \checkmark & - & \checkmark \\
min\_lateral\_signed             & - & \checkmark & - \\
min\_longit\_signed       & - & \checkmark & - \\
avg\_speed\_ego                   & \checkmark & - & \checkmark \\
speed\_at\_min\_dist              & - & \checkmark & \checkmark \\
max\_speed\_ego                   & \checkmark & \checkmark & \checkmark \\
max\_deceleration\_ego            & \checkmark & \checkmark & \checkmark \\
\bottomrule
\end{tabular}
\label{tab:interaction_features_usecases}
\end{table}

\subsection{Procedure and Metrics}

\head{\textbf{RQ\textsubscript{1}}}
We generated scenarios using both random sampling and NSGA-II search to obtain a balanced dataset of passing and failing scenarios for analyzing outcome consistency.
Next, we executed all scenarios in both \lofi and \hifi and labeled them with the corresponding \lofi and \hifi simulation outcomes; in total, we collected $592$ scenarios across all case studies, resulting in $1{,}184$ simulator executions.
Finally, we quantify outcome consistency between the simulators by comparing their labels scenario-wise, as follows:
(i)~Agreement on Pass (AP), when both simulators report a non-failure outcome;  
(ii)~Agreement on Fail (AF), when both simulators report a failure;  
(iii)~Disagreement Pass (DP), when the \hifi simulator reports a pass whereas the \lofi simulator reports a failure; and  
(iv)~Disagreement Fail (DF), otherwise.
AP and AF represent \textit{agreement} cases, while DP and DF correspond to \textit{disagreements} between simulators. We quantify consistency among simulators as the percentage of scenarios that fall into AF and AP within the evaluated set of scenarios. 

\head{\textbf{RQ\textsubscript{2}}/\textbf{RQ\textsubscript{3}}}
We used the dataset collected in RQ\textsubscript{1} to train binary classifiers to predict simulator agreement in the various scenarios. 
In particular, we used all scenario input parameters for training, i.e., we used 3 features (Pedestrian Speed, Ego Speed, Pedestrian Distance) for \autoware AEB, 4 features for \autoware CUT-IN, and 10 features for \frenetix.
To identify an appropriate classifier, we trained several models, including k-Nearest Neighbors (kNN), Random Forests (RF), Gradient Boosting (GBoost), and Support Vector Machines (SVM)~\cite{JodatSpurious23}. 
For each model, we optimized hyperparameters via grid search and evaluated performance using $5$-fold cross-validation (see \autoref{tab:model_performance}). 
%
For each use case, we set the threshold $th$ to $0.8$ based on the Youden index~\cite{Youden}, which maximizes the sum of sensitivity and specificity. This value was selected based on preliminary experiments; results for other thresholds are available in our replication package~\cite{replication-package}.
%



For RQ\textsubscript{2}, we measure the number of failures detected. We validate reported failures in \hifi simulation.
For RQ\textsubscript{3}, we measure the rate at which failures are discovered during the search process. Efficiency is the cumulative number of failures discovered over wall-clock time. 

To compare \toolname with the surrogate-assisted approach proposed by Jodat et al.~\cite{JodatSpurious23}, we trained a regression model to predict the fitness value for each generated scenario, using the same dataset collected for RQ\textsubscript{1}.
Specifically, we used only data from the \hifi runs, which serve as our ground truth, and split the data using the standard $80/20$ train–validation split for validation, as proposed in the original work.
We evaluated different dataset configurations, including balanced and imbalanced sets, and applied SMOTE as suggested in prior work~\cite{JodatSpurious23}. 
We evaluated several regression models, including Gaussian Processes (GP), Random Forest (RF), and a Multi-Layer Perceptron (MLP).
%
For each scenario, we used all input parameters as input features for the regression models and all fitness values as the models' target variables (\autoref{sec:objects}).

\begin{figure}[t]
    \centering
    \includegraphics[width=0.9\linewidth]{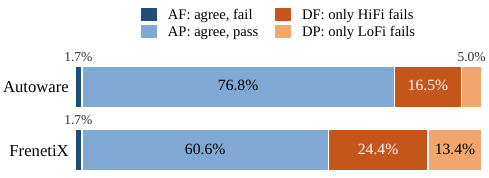}
\caption{RQ\textsubscript{1}: Distribution of simulator outcome agreement.
}
\label{fig:rq0-agreement-results}
\end{figure}

\head{\textbf{RQ\textsubscript{4}}}
We use for the scenario encoding the scenario-specific feature sets reported in \autoref{tab:interaction_features_usecases} and measured the diversity of failures using clustering~\cite{biagiola2023testing}.
First, we encode failing scenarios from RQ\textsubscript{2} across all testing approaches, as numerical feature vectors that capture key properties of the execution.
Next, we cluster the feature vectors using HDBSCAN~\cite{McInnes2017} with the Euclidean distance metric.
Finally, using the identified clusters, we compute the cluster coverage and the failure entropy, which quantifies the uniformity of failure distribution~\cite{sorokin2026stellar,biagiola2023testing}.
Based on preliminary experiments, we set the minimum cluster size to $3$ for \autoware and $6$ for \frenetix.

\head{Summary}
To address nondeterminism associated with search-based testing and improve the robustness of our conclusions, we repeated the experiments for \textbf{RQ\textsubscript{2}} and \textbf{RQ\textsubscript{3}} $n=6$ times.
Following established guidelines~\cite{Arcuri14statSE}, we assessed statistical significance using the Wilcoxon rank-sum test ($\alpha=0.05$) and effect sizes with Vargha--Delaney $\hat{A}_{12}$~\cite{vargha}.
In total, our experiments involved more than $27{,}000$ simulations, which correspond to approximately $480$ hours of computation.
All experiments were conducted on a dedicated machine equipped with an NVIDIA RTX 4090 GPU (24 GB VRAM), 32 GB RAM, and an Intel i7 16-core CPU. 

\subsection{Results}

\begin{table}[t]
\centering
\caption{RQ\textsubscript{1}: Agreement/disagreement classification performance used to instantiate \toolname. The best-performing model for each case study is highlighted in bold.}
\resizebox{\columnwidth}{!}{
\begin{tabular}{c l c c c}
\toprule
\textbf{Case Study} & \textbf{Model} & \textbf{Recall} & \textbf{Precision} & \textbf{F$_1$} \\
\midrule
\multirow{4}{*}{\autoware AEB} 
 & kNN    & 0.97 & 0.79 & \textbf{0.87} \\
 & RF     & 0.92 & 0.78 & 0.84 \\
 & GBoost & 0.78 & 0.79 & 0.79 \\
 & SVM    & 0.03 & 0.72 & 0.06 \\
\midrule
\multirow{4}{*}{Autoware CUT-IN}
 & kNN    & 0.81 & 0.83 & 0.81 \\
 & RF     & 0.91 & 0.81 & \textbf{0.86} \\
 & GBoost & 0.93 & 0.80 & \textbf{0.86} \\
 & SVM    & 0.95 & 0.77 & 0.85 \\
\midrule
\multirow{4}{*}{\frenetix} 
 & kNN    & 0.67 & 0.73 & 0.70 \\
 & RF     & 0.76 & 0.74 & \textbf{0.75} \\
 & GBoost & 0.69 & 0.74 & 0.72 \\
 & SVM    & 0.54 & 0.82 & 0.65 \\
\bottomrule
\end{tabular}
}
\label{tab:model_performance}
\end{table}


\input{content/table-rq2}

\subsubsection{Agreement (RQ\textsubscript{1})}

\autoref{fig:rq0-agreement-results} summarizes the distribution of agreement and disagreement across all generated scenarios.
Overall, the simulators agree on $78.5\%$ of scenarios for \autoware and $62.3\%$ for \frenetix.

Most agreement corresponds to AP, whereas AF is comparatively rare, accounting for only $1.7\%$ of all scenarios.
Among disagreement cases, DF is substantially more frequent than DP, indicating that \lofi more often misses failures detected by \hifi than the converse.
These results indicate that simulator agreement is concentrated on non-failing scenarios, whereas disagreement primarily occurs in safety-critical situations.\footnote{Considering failure as the positive class, failure precision and recall are computed as $\mathrm{Precision}=AF/(AF+DP)$ and $\mathrm{Recall}=AF/(AF+DF)$. For \autoware, precision/recall are $6.5\%/11.3\%$; for \frenetix, they are $9.3\%/25.4\%$. Considering pass as the positive class, precision and recall are $\mathrm{Precision}=AP/(AP+DF)$ and $\mathrm{Recall}=AP/(AP+DP)$, yielding $81.9\%/71.3\%$ for \autoware and $93.9\%/82.3\%$ for \frenetix.}
This behavior is consistent with the fidelity gap between the simulators and motivates our orchestration strategy, which reserves computationally intensive \hifi execution for scenarios where increased simulation fidelity influences the outcome.

\autoref{tab:model_performance} reports the classification performance.
The best-performing models were kNN ($k=7$) for \autoware AEB, Gradient Boosting for \autoware CUT-IN, and Random Forest for \frenetix, achieving F$_1$ scores of $0.87$, $0.86$, and $0.75$, respectively.
We use these classifiers to instantiate \toolname in the remaining research questions, demonstrating that simulator agreement can be predicted with sufficient accuracy to support dynamic simulator orchestration.

These findings indicate that disagreement occupies a relatively small but safety-critical portion of the scenario space. Consequently, simulator orchestration can substantially reduce HiFi executions while preserving access to the scenarios where increased fidelity is most valuable.

\begin{tcolorbox}[boxrule=0pt,frame hidden,sharp corners,enhanced,borderline north={1pt}{0pt}{black},borderline south={1pt}{0pt}{black},boxsep=2pt,left=2pt,right=2pt,top=2.5pt,bottom=2pt]
\textbf{RQ\textsubscript{1} (agreement).}
\textit{\lofi and \hifi agree on $78.5\%$ of scenarios for \autoware and $62.3\%$ for \frenetix. Agreement is concentrated on non-failing scenarios and can be accurately predicted from scenario parameters ($F_1 > 0.75$), providing the foundation for simulator orchestration.}
\end{tcolorbox}

\subsubsection{Effectiveness (RQ\textsubscript{2})}

\autoref{tab:combined_results_side_by_side_with_surrogate} reports for each of the three scenarios, six runs, and four approaches (the two single-fidelity baselines, the surrogate-assisted baseline, and \toolname), the number of generated (col. \emph{Tests}) and failed (col. \emph{Fail.}) tests, the pass/fail ratio (col. \emph{Ratio}), and the average test execution time (col. \emph{Time/Test}).
For each run, Time/Test is the search budget in seconds divided by the number of evaluated candidates. This metric includes search and orchestration overhead but excludes subsequent HiFi failure validation.
 The table also reports results aggregated across the runs (rows \emph{Avg}) and highlights statistically significant improvements. 

We observe that \lofi executes the largest number of tests across all configurations but identifies the fewest failures. In contrast, \hifi executes substantially fewer tests because of its higher computational cost, yet detects more failures than \lofi. These findings are consistent with the expected trade-off between the two simulation fidelities: \lofi enables efficient exploration of the scenario space but has limited failure detection capability due to its lower fidelity, whereas \hifi provides more effective failure identification at the expense of reduced exploration.
The surrogate-assisted baseline generates, on average, a similar number of tests as \lofi; however, it identifies up to 12$\times$ more failures than \lofi. Compared to \hifi, it identifies a higher number of failures in two out of three case studies, i.e., \frenetix and \autoware AEB.

These results highlight the benefits of using a surrogate model to balance execution cost and failure identification. However, for each case study, our approach \toolname achieves the highest number of failures: for the \autoware AEB case study, \toolname identifies up to 50\% more failures than surrogate-based search, while for \autoware CUT-IN it finds up to 37\% more failures than \hifi.  For \frenetix, it achieves up to 20\% more failures than the baselines.
On average, \toolname identifies up to 2.1$\times$ more failures than single simulator-based testing in \hifi.
All differences between \toolname and other approaches are statistically significant ($p < 0.05$) with large effect sizes. The complete table of statistical test results is provided in the replication package~\cite{replication-package}.

\head{Classifier Accuracy} To analyse the classifiers performance on predicting simulator agreements and disagreement during search, we audited 10 randomly sampled candidates per configuration, distributed across the six search runs and different search stages, including both LoFi- and HiFi-routed scenarios. We executed each candidate in the simulator for which its outcome was missing and compared the resulting agreement label with the classifier’s prediction made during search.
Precision/recall/F1 during search are .76/.72/.74 for AEB and .77/.74/.76 for FrenetiX, compared with .79/.97/.87 and .74/.76/.75, respectively. The classifier maintains similar precision in both settings and comparable overall performance for FrenetiX; AEB recall decreases, yielding an F1 of .74 during search. CUT-IN study results are provided in the replication package~\cite{replication-package}.

\head{Ablation Study}
To isolate the contribution of the orchestration mechanism, we replaced the classifier-guided simulator selection with random LoFi/HiFi selection while keeping all remaining components of SimFusion unchanged. This ablation evaluates whether the observed improvements stem from informed simulator selection rather than simply combining two simulators.
Across six independent runs, \toolname improved failure discovery by 17\% for \autoware and 22\% for \frenetix compared to random orchestration. These differences were statistically significant, indicating that the observed improvements stem from informed simulator orchestration rather than simply alternating between \lofi and \hifi simulations. These results demonstrate that the performance gains originate from learning where HiFi execution is beneficial, rather than from the availability of multiple simulators alone.

\head{Missed failures} In addition, we rerun a sample of tests across all seeds that have been not executed in \hifi as they passed in \lofi. For the AEB case study, 20\% of the sampled cases were missed failures, while for \frenetix 16\% were. Results for the Cutin case study are available in the replication package~\cite{replication-package}.

\begin{tcolorbox}[boxrule=0pt,frame hidden,sharp corners,enhanced,borderline north={1pt}{0pt}{black},borderline south={1pt}{0pt}{black},boxsep=2pt,left=2pt,right=2pt,top=2.5pt,bottom=2pt]
\textbf{RQ\textsubscript{2} (effectiveness).}
\textit{Compared to single-fidelity and surrogate assisted testing,  \toolname substantially improves failure detection by finding, on average, $2.1\times$ more failures than \hifi and $1.6\times$ more than surrogate-assisted testing across the three case studies.}
\end{tcolorbox}

\begin{figure}
    \centering
    \includegraphics[width=\linewidth]{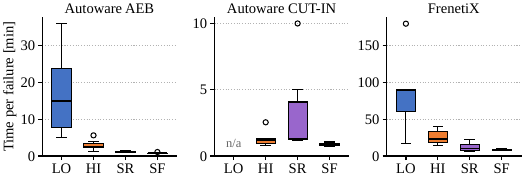}
    \caption{RQ\textsubscript{3}: Time per failure for \lofi (LO), \hifi (HI), surrogate-based testing (SR), and \toolname (SF).}
    \label{fig:efficiency-time-per-failure}
\end{figure}

\subsubsection{Efficiency (RQ\textsubscript{3})}

\autoref{fig:efficiency-time-per-failure} reports the average time required to detect a failure. 
Across all case studies, \toolname consistently achieves the shortest detection time. The best result is obtained for the \autoware CUT-IN scenario (0.86 min), compared to 1.3 min for \hifi. \frenetix is the most challenging case study for all approaches; nevertheless, \toolname detects failures in 8.30 min on average, 30\% faster than the surrogate-based approach. In addition, \toolname consistently exhibits the lowest runtime variability across all scenarios.
On average across all scenarios, \toolname identifies up to 15.85$\times$ faster failures than LoFi (52.32 minutes). Compared to \hifi, \toolname is up to 3$\times$ faster than HiFi (9.93 minutes), and 1.7 $\times$ faster than the surrogate-based approach (5.54 minutes). 
All differences between \toolname and other approaches are statistically significant with large effect sizes~\cite{replication-package}.

To take into account the time required during the training phase to train the classifier, collecting the paired LoFi/HiFi training data requires, for instance, 7.4 hours for the Autoware case study. 
The predictor is reused across all six search runs. Including this one-time cost, the average time per validated failure increases from 44s to 62s, remaining 37\% lower than surrogate-assisted testing (99s including surrogate model training). This comparison reflects six campaigns sharing the initialization cost; its contribution per campaign is amortized with further reuse.

We evaluate the number of saved \hifi executions in \toolname. \autoref{tab:hifi_execution_counts} shows that on average up to 31.4\% fewer \hifi tests are executed with \toolname than with \hifi only test generation.

\begin{table}[t]
    \centering
    \caption{HiFi execution counts. The table reports the number of SimFusion's HiFi executions vs. HiFi-only tests.}
    \label{tab:hifi_execution_counts}
    \small
    \setlength{\tabcolsep}{5pt}
    \begin{tabular}{lrrr}
        \toprule
        \textbf{Case study}
        & \textbf{\#SimFusion HiFi}
        & \textbf{\#HiFi}
        & \textbf{Reduction} \\
        \midrule
        Autoware CUT-IN 
        & 1,313 & 1,920 & 31.6\% \\

        Autoware AEB 
        & 1,447 & 2,160 & 33.0\% \\

        FrenetiX
        & 325 & 420 & 22.6\% \\
        \midrule
        \textbf{Total}
        & \textbf{3,085}
        & \textbf{4,500}
        & \textbf{31.4\%} \\
        \bottomrule
    \end{tabular}
\end{table}

\begin{tcolorbox}[boxrule=0pt,frame hidden,sharp corners,enhanced,borderline north={1pt}{0pt}{black},borderline south={1pt}{0pt}{black},boxsep=2pt,left=2pt,right=2pt,top=2.5pt,bottom=2pt]
\textbf{RQ\textsubscript{3} (efficiency).}
\textit{\toolname discovers failures faster than baseline approaches. \toolname finds on average up to 3$\times$ faster failures than \hifi and up to 1.7$\times$ faster than surrogate-assisted testing.}
\end{tcolorbox}

\subsubsection{Diversity (RQ\textsubscript{4})}

The entropy and coverage results for the evaluation of the failure diversity are provided in \autoref{tab:diversity_results_switched}. 

We can observe that \toolname achieves for both \autoware case studies the highest coverage of 94\% and 100\%, substantially outperforming \hifi with 60\% and 83\%, \lofi with 23\% and 0$\%$, and the surrogate baseline with 11\% and 17\%.
For \autoware CUT-IN \toolname achieves the highest entropy, while for the \autoware AEB scenario the surrogate-based approach achieves a higher value.

For \frenetix, both \toolname and the surrogate baseline achieve higher failure coverage and entropy than the single-fidelity baselines. \toolname achieves the highest coverage overall ($0.88$) and the highest entropy ($0.58$), indicating that it explores a broader set of failure behaviors.

These results indicate that SimFusion explores previously unseen regions of the scenario space rather than repeatedly identifying similar failure instances.




\begin{tcolorbox}[boxrule=0pt,frame hidden,sharp corners,enhanced,borderline north={1pt}{0pt}{black},borderline south={1pt}{0pt}{black},boxsep=2pt,left=2pt,right=2pt,top=2.5pt,bottom=2pt]
\textbf{RQ\textsubscript{4} (diversity).}
\textit{\toolname achieves the highest failure diversity overall, with an average failure coverage of $94\%$ and an average entropy of 0.45 across all scenarios. In the majority of cases, \toolname outperforms all the baseline approaches regarding failure diversity.}
\end{tcolorbox}

\begin{table}[t]
\centering
\caption{RQ\textsubscript{4}: Failure coverage and entropy (in \%; higher is better). Best values per column in bold.}
\label{tab:diversity_results_switched}
\small
\setlength{\tabcolsep}{3pt}
\begin{tabular}{l *{6}{>{\centering\arraybackslash}p{0.9cm}}}
\toprule
& \multicolumn{2}{c}{\footnotesize\textbf{\autoware AEB}}
& \multicolumn{2}{c}{\footnotesize\textbf{\autoware CUT-IN}}
& \multicolumn{2}{c}{\footnotesize\textbf{\frenetix}} \\
\cmidrule(lr){2-3} \cmidrule(lr){4-5} \cmidrule(lr){6-7}
\textbf{Approach} & Cov. & Ent. & Cov. & Ent. & Cov. & Ent. \\
\midrule
\lofi        & 23 & 64 & 0 & 0 & 22 & 31 \\
\hifi        & 60 & 80 & 83 & 32 & 77 & 48 \\
Surrogate    & 11 & \textbf{91} & 17 & 7 & 55 & 54 \\
\toolname    & \textbf{94} & 43 & \textbf{100} & \textbf{34} & \textbf{88} & \textbf{58} \\
\bottomrule
\end{tabular}
\end{table}

\section{Threats To Validity}
\label{sec:threats}


\head{Internal Validity}
Differences in simulator implementations may affect scenario alignment and execution outcomes. 
For \autoware, we verified the alignment of pedestrian trajectories between the \lofi and \hifi simulators and selected the CARLA step size based on preliminary experiments ($0.15$\,s), while otherwise using default configurations. 
For \frenetix, we rely on the MultiDrive framework~\cite{kaufeld2025multidrive}, which ensures consistent scenario execution across the \lofi and \hifi simulators. 
Since \lofi outcomes may be unreliable (as shown in RQ\textsubscript{1}), all failures by all approaches in \lofi were validated in \hifi. 

Both simulators use the same fitness definitions, units, and optimization directions, and their values are compared directly during selection. 
Our paired minimum-distance analysis shows closer numerical agreement for jointly failing scenarios than for jointly passing scenarios (see \cite{replication-package}). However, differences can still affect candidate rankings.
%
Implementation errors could also affect the results. To mitigate this threat, we built \toolname on top of the OpenSBT framework~\cite{sorokin2023opensbt}, which provides a modular and well-tested infrastructure for search-based testing of ADS.

\head{Construct Validity}
Simulator flakiness could influence the observed failures. To mitigate this threat, we re-executed scenarios in the high-fidelity simulators whenever inconsistent outcomes occurred. Preliminary experiments did not indicate substantial flakiness. 
Another threat concerns whether our metrics accurately capture the effectiveness and diversity of test generation. To address this, we rely on widely used metrics for failure discovery and diversity analysis, and we repeated each experiment multiple times using randomized search-based algorithms and assessed statistical significance. 


\head{External Validity}
Our evaluation considers one LoFi/HiFi pair per ADS stack. Although the orchestration strategy is independent of the underlying simulators, additional studies involving different simulator pairs and fidelity gaps would further assess the generality of the proposed approach.
Our results may depend on the specific ADS stacks, simulators, and scenarios used in our evaluation. To mitigate this threat, we evaluated \toolname on two distinct ADS stacks (\autoware and \frenetix) across four simulators with varying fidelity levels and simulation paradigms, as well as three different abstract scenarios.

%% file: content/table-rq2.tex

\begin{table*}[t]
\centering

\resizebox{\textwidth}{!}{

\begin{threeparttable}
\caption{RQ\textsubscript{2}: Effectiveness results for three case studies. Failure count and ratio are computed for validated failures (in \hifi). Runtime/test denotes the average runtime per executed test case (s). Best results across methods are highlighted in bold.}
\label{tab:combined_results_side_by_side_with_surrogate}

\footnotesize
\setlength{\tabcolsep}{2.5pt}
\renewcommand{\arraystretch}{1}

\begin{tabular}{lrrrrrrrrrrrrrrrrrr}
\toprule
& Run
& \multicolumn{4}{c}{\textbf{LoFi}}
& \multicolumn{4}{c}{\textbf{HiFi}}
& \multicolumn{4}{c}{\textbf{Surrogate}}
& \multicolumn{4}{c}{\textbf{SimFusion}} \\
\cmidrule(lr){3-6} \cmidrule(lr){7-10} \cmidrule(lr){11-14} \cmidrule(lr){15-18}
 & & Tests  ↑  & Fail. ↑ & Ratio ↑ & Time/Test ↓
   & Tests & Fail. & Ratio & Time/Test
   & Tests & Fail. & Ratio & Time/Test
   & Tests & Fail. & Ratio & Time/Test \\
\midrule

\multirow{7}{*}{\autoware AEB}
 & Run 1  & 660 & 10  & 0.01 & 16.36 & 340 & 69  & 0.20 & 31.76 & 280 & 181 & 0.65 & 38.57 & 378 & 289 & 0.76 & 28.57 \\
 & Run 2  & 680 & 5   & 0.00 & 15.88 & 320 & 32  & 0.10 & 33.75 & 300 & 185 & 0.62 & 36.00 & 440 & 257 & 0.58 & 24.55 \\
 & Run 3  & 620 & 35  & 0.05 & 17.42 & 380 & 150 & 0.39 & 28.42 & 400 & 124 & 0.31 & 27.00 & 460 & 254 & 0.55 & 23.48 \\
 & Run 4  & 680 & 7   & 0.01 & 15.88 & 380 & 74  & 0.19 & 28.42 & 320 & 146 & 0.45 & 33.75 & 420 & 166 & 0.39 & 25.71 \\
 & Run 5  & 680 & 15  & 0.02 & 15.88 & 360 & 85  & 0.23 & 30.00 & 280 & 198 & 0.71 & 38.57 & 380 & 234 & 0.61 & 28.42 \\
 & Run 6  & 660 & 28  & 0.04 & 16.36 & 380 & 47  & 0.12 & 28.42 & 280 & 150 & 0.53 & 38.57 & 380 & 281 & 0.73 & 28.42 \\
\cmidrule(lr){2-18}
 & Avg. & 663.33 & 16.66 & 0.02 & 16.30 & 360.00 & 76.16 & 0.20 & 30.13 & 310.00 & 164.00 & 0.55 & 34.84 & 409.66 & \textbf{246.83}\tnote{*} & \textbf{0.60} & 26.53 \\
 \midrule

\multirow{7}{*}{\autoware CUT-IN}
& Run 1 & 615 & 0 & 0 & 14.6 & 330 & 135 & 0.41 & 27.27 & 930  & 123 & 0.13 &  9.68 & 315 & 197 & 0.62 & 28.57 \\
& Run 2 & 630 & 0 & 0 & 14.3 & 375 & 120 & 0.32 & 24.00 & 705  & 30  & 0.04 & 12.77 & 315 & 148 & 0.47 & 28.57 \\
& Run 3 & 630 & 0 & 0 & 14.3 & 345 & 182 & 0.53 & 26.09 & 300  & 15  & 0.05 & 30.00 & 285 & 181 & 0.64 & 31.57 \\
& Run 4 & 600 & 0 & 0 & 15  & 300 & 165 & 0.55 & 30.00 & 570  & 125 & 0.21 & 15.79 & 315 & 215 & 0.68 & 28.57 \\
& Run 5 & 600 & 0 & 0 & 15 & 270 & 59  & 0.22 & 33.33 & 1290 & 117 & 0.09 &  6.98 & 270 & 181 & 0.67 & 33.33 \\
& Run 6 & 600 & 0 & 0 & 15 & 300 & 111 & 0.39 & 30.00 & 1155 & 116 & 0.10 &  7.79 & 330 & 142 & 0.43 & 27.29 \\
\cmidrule(lr){2-18}
& Avg. & 612.5 & 0 & 0 & 14.7 & 320.00 & 128.67 & 0.40 & 28.45 & 825.00 & 87.67  & 0.10 & 13.83
     & 305 & \textbf{177.33}\tnote{*} & \textbf{0.58} & 29.65 \\
\midrule

\multirow{7}{*}{\frenetix}
 & Run 1 & 258 & 4  & 0.01 & 83.72 & 90 & 26 & 0.28 & 240.00 & 140 & 16 & 0.11 & 154.29 & 110 & 37 & 0.34 & 196.36 \\
 & Run 2 & 260 & 6  & 0.02 & 83.08 & 90 & 20 & 0.22 & 240.00 & 370 & 22 & 0.06 & 58.38 & 110 & 47 & 0.43 & 196.36 \\
 & Run 3 & 258 & 4  & 0.01 & 83.72 & 60 & 10 & 0.16 & 360.00 & 100 & 28 & 0.25 & 216.00 & 109 & 45 & 0.41 & 198.17 \\
 & Run 4 & 260 & 2  & 0.00 & 83.08 & 60 & 9  & 0.15 & 360.00 & 350 & 47 & 0.13 & 61.71 & 120 & 39 & 0.32 & 180.00 \\
 & Run 5 & 290 & 0  & 0.00 & 74.48 & 60 & 17 & 0.28 & 360.00 & 290 & 50 & 0.17 & 74.48 & 130 & 48 & 0.37 & 166.15 \\
 & Run 6 & 310 & 21 & 0.06 & 69.67 & 60 & 15 & 0.25 & 360.00 & 350 & 56 & 0.16 & 61.71 & 130 & 47 & 0.36 & 166.15 \\
\cmidrule(lr){2-18}
 & Avg. & 272.66 & 6.16 & 0.02 & 79.63 & 70.00 & 16.16 & 0.22 & 320.00 & 266.67 & 36.50 & 0.15 & 81.00 & 118.17 & \textbf{43.83}\tnote{*} & \textbf{0.37} & 183.86 \\

\midrule

\multicolumn{2}{l}{Avg.}
 & 516.16 & 7.61 & 0.01 & 36.88
 & 250.00 & 73.66 & 0.27 & 126.19
 & 467.22 & 96.06 & 0.27 & 43.22
 & 277.61 & \textbf{155.99}\tnote{*} & \textbf{0.52} & 80.01 \\
\bottomrule
\end{tabular}
\begin{tablenotes}[flushleft]
\footnotesize
\item[*] Statistically significant improvement over the best baseline (Wilcoxon rank-sum test, $\alpha=0.05$) with large effect size (Vargha--Delaney $\hat{A}_{12}$).
\end{tablenotes}
\end{threeparttable}

}

\end{table*}

%% file: content/discussion.tex
\section{Discussion}
\label{sec:discussion}

\head{Applicability}
The effectiveness of SimFusion depends primarily on the relationship between the two simulation fidelities rather than on the specific simulators employed. In particular, SimFusion is expected to provide the greatest benefit when LoFi and HiFi agree on the majority of scenarios, while disagreements remain concentrated in a relatively small subset of safety-critical situations. Under these conditions, inexpensive LoFi executions can safely filter most candidate scenarios, allowing computationally intensive HiFi simulation to be reserved for cases where increased realism is likely to influence the testing outcome.

Applying SimFusion to additional simulator pairs also requires that equivalent scenarios can be executed consistently across both environments. As simulators inevitably differ in map representations, vehicle dynamics, and timing models, validating scenario alignment is essential to ensure that observed disagreements arise from differences in simulation fidelity rather than inconsistencies in scenario execution. In our case studies, we verified alignment using trajectory similarity metrics, although alternative validation strategies may be appropriate depending on the simulation platforms.

\head{Predicting Simulator Agreement vs. Approximating Simulation}
Recent surrogate-assisted testing approaches reduce simulation cost by learning mappings from scenario parameters to fitness values or directly to test outcomes~\cite{JodatSpurious23,DBLP:conf/wcre/BirchlerGKGP22}. Such approaches implicitly attempt to approximate the behavior of the complete autonomous driving system, whose outputs emerge from complex interactions among perception, planning, vehicle dynamics, control, and the environment.

SimFusion adopts a different perspective. Rather than approximating simulator outputs, it predicts whether two simulators are likely to agree on a scenario's outcome. Our results suggest that learning simulator agreement is a substantially simpler classification problem than approximating ADS behavior itself, while still providing sufficient information to guide simulator orchestration. We believe this represents a complementary direction for reducing simulation cost in search-based testing of ADS.

\head{Understanding Simulators Disagreement}
Our empirical analysis shows that disagreements between LoFi and HiFi are not uniformly distributed across the scenario space. Instead, they are concentrated in safety-critical situations, whereas the vast majority of nominal scenarios are classified consistently by both simulators. This behavior is consistent with the abstractions introduced by low-fidelity simulation, which simplify sensing, perception, and vehicle dynamics while largely preserving nominal driving behavior.

This asymmetry explains why multi-fidelity orchestration is effective. Executing every scenario in HiFi provides little additional information for the large fraction of scenarios where both simulators agree. Conversely, selectively invoking HiFi in regions where disagreement is likely enables computational resources to be focused where increased simulation fidelity has the greatest influence on testing outcomes.

\head{Scenario Complexity}
The benefits of SimFusion become more pronounced as the complexity of the interaction space increases. While the Autoware case studies involve relatively constrained interactions, the FrenetiX scenario contains multiple vehicles, lane changes, and richer traffic dynamics, resulting in a substantially larger space of possible behaviors. In such settings, rapid exploration through LoFi simulation allows considerably more candidate scenarios to be evaluated within a fixed testing budget, while selective HiFi validation preserves the ability to identify fidelity-dependent failures.

Another perspective is that each simulator induces a different search landscape over the same scenario space. Search guided exclusively by a single simulator explores failures exposed under that simulator's abstractions~\cite{matteoMaxibon2024}. In contrast, SimFusion exploits both landscapes by concentrating computational effort on regions where simulator disagreement is most likely to occur. Similar ideas have recently been explored by combining multiple high-fidelity simulators to expose simulator-dependent failures~\cite{multisim2026sorokin,matteoMaxibon2024}. Our work shows that analogous benefits can also be achieved by orchestrating simulators across different fidelity levels.





%% file: content/related.tex
\section{Related Work}\label{sec:related}


\noindent\textbf{Single-simulator testing approaches} generate scenarios automatically via search-based software testing or fuzzing~\cite{survey-lei-ma}.
Most current test generation methods use search-based approaches to automatically create test scenarios for DNN-based ADS~\cite{moghadam2023machine,2020-Riccio-FSE,Abdessalem-ASE18-1,Abdessalem-ASE18-2,Abdessalem-ICSE18,deepxplore,deeptest,deeproad,Kim:2019:GDL:3339505.3339634}.  
In these approaches, test generators vary scenario attributes, including road topologies~\cite{asfault,2020-Riccio-FSE}, traffic configurations~\cite{liETAL2020}, road users~\cite{Abdessalem-ASE18-1,Abdessalem-ASE18-2,Abdessalem-ICSE18}, and environmental elements~\cite{Sun-et_al26a}.
Relevant work in fuzzing includes DriveFuzz~\cite{drivefuzz}, which leverages the vehicle's physical state and oracles grounded in real-world traffic rules to guide the fuzzer in uncovering potential misbehaviors; AutoFuzz~\cite{zhongETAL2021}, which focuses on fuzzing the test scenario specifications and employs a binary classifier to identify highly critical scenario seeds; and AV-Fuzzer~\cite{liETAL2020}, which extends a genetic algorithm with information about the positioning of globally monitored non-player characters in each driving scenario. 
%
%
Cheng et al.~\cite{10.1145/3597926.3598072} introduce BehaviorMiner, an unsupervised model that extracts temporal features from predefined scenarios and employs clustering-based abstraction to group behaviors with similar features into abstract states. 
Because these approaches rely on a single simulator, their cost-effectiveness depends strongly on the simulator's fidelity, necessitating a trade-off between computational time (for higher-fidelity simulations) and reliability of the results~\cite{8814268}.
In contrast, our work leverages multiple simulators with different fidelity levels, enabling more efficient exploration while preserving accurate failure assessment.

\noindent\textbf{Multi-simulator testing approaches} use multiple simulators to improve test reliability.
Borg et al.~\cite{digitalTwinAreNotMonozygotic} and Amini et al.~\cite{AminiFlaky2024} showed that relying on a single simulator yields unreliable conclusions about the quality of the system under test because some failures stem from simulator-specific characteristics rather than issues in the SUT.
To address this issue, Biagiola et al.~\cite{matteoMaxibon2024} and Sorokin et al.~\cite{multisim2026sorokin} used ensembles of simulators to identify and resolve discrepancies in scenario executions across simulators. 
However, SimFusion addresses a different objective: improving testing efficiency relative to a designated HiFi reference. It predicts LoFi/HiFi agreement to avoid unnecessary HiFi executions.
%
More recently, Kaufeld et al.~\cite{kaufeld2025multidrive} developed a co-simulation framework that integrates a low-fidelity and a high-fidelity simulator to facilitate scenario transfer and analysis across simulators of different fidelity levels.
MFBO-Drive~\cite{DBLP:conf/ivs/OsikowiczMWS26} applies multi-fidelity Bayesian optimization by adaptively selecting both candidate scenarios and the fidelity level at which they are evaluated, reducing computational cost while maintaining test quality. In contrast, our approach orchestrates different simulators based on predicted cross-fidelity agreement. Compared to these approaches, our work improves testing efficiency by combining low- and high-fidelity simulators to limit the execution of expensive simulations while preserving ADS testing reliability.

\noindent\textbf{Surrogate-based scenario generation} improves testing efficiency by guiding scenario generation using computationally efficient approximations based on machine learning or probabilistic models.
Mullins et al.~\cite{MULLINS2018197} employed Gaussian Processes to steer the search toward unexplored regions of the input space, thereby increasing coverage. 
Biagiola and Tonella~\cite{biagiola2023testing} trained a classifier on past execution data to guide failure discovery in Deep Reinforcement Learning systems.
Ul Haq et al.~\cite{DBLP:conf/icse/Haq0B22} extended many-objective search to utilize surrogate models that mimic simulators' output.
Jodat et al.~\cite{JodatSpurious23} proposed learning interpretable failure models that approximate test outcomes using multiple surrogate predictors. 
These approaches reduce testing costs by approximating simulator executions through predictive models. In contrast, \toolname leverages existing simulators with different fidelity levels and dynamically orchestrates their use during testing.

%% file: content/conclusion.tex
\section{Conclusions and Future Work}
\label{sec:conclusion}

We presented \toolname, a multi-fidelity search-based testing framework that improves the effectiveness and efficiency of search-based scenario testing by orchestrating simulators across multiple fidelity levels. Using a classifier that predicts cross-simulator agreement, \toolname selectively executes scenarios in \lofi or \hifi simulation, combining efficient exploration with reliable failure detection.
Our evaluation across two case studies, four driving simulators, and three abstract scenarios shows that using simulation agreement to guide fidelity selection identifies more failures, with higher failure coverage, than single-fidelity and surrogate-assisted approaches.

As future work, we plan to investigate how \toolname can be extended to leverage multiple parallel \lofi simulators combined with multiple parallel \hifi simulators. Such configurations could further accelerate exploration of the scenario space while maintaining reliable failure validation. More broadly, we aim to study how adaptive multi-fidelity simulation pipelines can support continuous testing and faster development cycles for safety-critical autonomous systems.
